\documentclass[pre,aps,reprint,showpacs,superscriptaddress]{revtex4-1}
\usepackage{graphicx}
\usepackage{amsmath}
\usepackage{dcolumn}
\usepackage{bm}
\usepackage{lineno}
\usepackage{color}
\begin{document}


\title{Magetostatic amplifier with tunable maximum by twisted-light plasma interactions}

\author{D. Wu}
\email{Email:\ wudong@siom.ac.cn}
\affiliation{State Key Laboratory of High Field Laser Physics, 
Shanghai Institute of Optics and Fine Mechanics, 201800 Shanghai, China}
\affiliation{Helmholtz Institut Jena, D-07743 Jena, Germany}
\author{J. W. Wang}
\email{Email:\ j.wang@gsi.de}
\affiliation{Helmholtz Institut Jena, D-07743 Jena, Germany}
\affiliation{State Key Laboratory of High Field Laser Physics, 
Shanghai Institute of Optics and Fine Mechanics, 201800 Shanghai, China}

\date{\today}

\begin{abstract}
Laser beams with Laguerre-Gaussian (LG) mode carry orbital angular momentum (OAM), however when interacting with plasmas, the net OAM acquired by plasmas is basically zero after interaction. Here, we find when there exist a small magetostatic seed along laser propagation direction, the barrier would be broken, giving rise to dramatic OAM transfer from LG lasers to plasmas. Hence, the net OAM remained in plasmas system would continuously enhance the magetostatic field, until the corresponding Larmor frequency of electrons is comparable to the laser frequency in vacuum. 
Three-dimensional particle-in-cell (3D-PIC) simulations are performed to confirm our theory, producing space-uniform, time-stable and extremely intense magetostatic fields.
\end{abstract}

\pacs{52.38.Kd, 41.75.Jv, 52.35.Mw, 52.59.-f}

\maketitle

\textit{\textbf{Introduction}}--Generation of space-uniform, time-stable and extremely-strong magnetostatic fields will be of great importance to plasma, beam physics and astrophysics \cite{sm1,sm2,sm3,sm4,sm5,sm6,sm7,sm8,sm9}. 
Among many of great applications, such as the magnetically-assisted fast ignition approach \cite{sm8}, the axial magnetic field would collimate the relativistic electron beam to increase the coupling efficiency between the laser and the core. Furthermore, the magnetically-assisted proton acceleration scheme was also proposed recently \cite{sm9}, the axial magnetic field could not only collimate the relativistic electron beam but also enhance the electron heating efficiency in front of the target. Although several kilotesla field has been measured in a relativistically intense laser-plasma interaction experiment \cite{m1,m2}, it is in small spatial and temporal scales, which is difficult in applications. Thanks to the landmark work performed recently \cite{m3}, extremely strong magetostatic field up to 
$1500$ T is obtained by using a novel capacitor-coil target design making use of the hot electrons generated during the laser-target interactions. However even so, this state-of-art strongest magetostatic field is still far smaller when comparing with laser field. 

Importantly, the schemes to produce relativistic twisted-light (with intensity over $10^{18}\ \text{W}/\text{cm}^2$) have been recently proposed \cite{lg1,lg2} and initiated new investigations of twisted-laser at relativistic intensities \cite{lg2,lg3,lg4,lg5,lg6,lg7}. 
Prompted by the inspiring advances of the twisted light intensity, 
the behaviour of twisted-laser-plasma interactions at relativistic intensities should be studied in detail. 
Relativistic intense twisted-light carries huge orbital angular momentum (OAM), once transferred to plasmas in the interaction, one can imagine, extremely strong axial magetostatic field could be generated and sustained by the cyclotron motion of plasmas.  

In this paper, we have investigated relativistic intense twisted-light plasma interactions basically in the first-order approximation. To first order, plasma is assumed to be of uniform, the longitudinal component of laser field is ignored when compared with the transverse parts. Furthermore, instabilities \cite{he2,he3}, such as filament instabilities, are ignored in the considered situations. We find that, when twisted light interacts with plasmas, the net OAM acquired by plasmas is basically zero after interaction. However, we also find when imposing a small magetostatic seed along laser propagation direction, the barrier would be broken, giving rise to dramatic OAM transfer from LG lasers to plasmas. Hence, the net OAM remained in plasmas system would continuously enhance the magetostatic field, until the corresponding Larmor frequency ($\omega_c\equiv eB/m_e$) of electrons is comparable to the laser frequency ($\omega_0$) in vacuum. Three-dimensional particle-in-cell (3D-PIC) simulations are performed to confirm our theory, producing space-uniform, time-stable and extremely intense magetostatic fields.

\begin{figure}
\includegraphics[width=8.5cm]{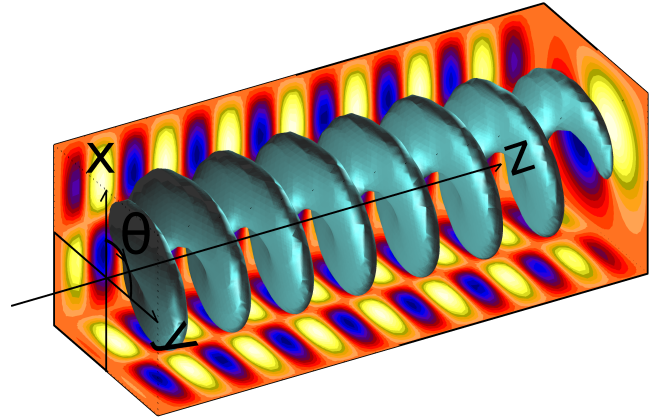}
\caption{\label{fig1} (color online) Schematic of twisted light ($l=1$ and $p=0$) propagating in z direction. The twisted ``bulk'' is the surface of constant $E_x$. The x-y, x-z and y-z cross-sections of $E_x$ are also presented at the corresponding walls.}
\end{figure}

\textit{\textbf{Theoretical model}}--We consider a laser pulse carrying OAM, whose spatial profile will be expressed as an LG mode. Considering the wave propagating along z-direction, these LG modes reads,
\begin{eqnarray}
\text{LG}_{l,p}(r,\ \theta,\ z) &=&\sqrt{\frac{2p!}{\pi(\lvert l\rvert+p)!}}\times\frac{b_0}{b}\times\text{L}_{p}^{\lvert l\rvert}(\frac{2r^2}{b^2})\times(\frac{\sqrt{2}r}{b})^{\lvert l\rvert} \nonumber \\ 
&\times& \exp(-\frac{r^2}{b^2})\times\exp(\text{i}l\theta+\text{i}\frac{k_0r^2}{2R}+\text{i}\Phi), 
\end{eqnarray}
where $r=\sqrt{x^2+y^2}$, $\theta=\arctan(y/x)$, $b\equiv b(z)=b_0\sqrt{1+(z/z_0)^2}$ is the beam width, $b_0$ is the beam waist at $z=0$, $z_0=k_0b^2_0/2$ is the Rayleigh range, $k_0$ is the carrier wave number, $R\equiv R(z)=z[1+(z_0/z)^2]$ is the phase-front radius, $\Phi\equiv\Phi(z)=-(2p+\lvert l\rvert+1)\arctan(z/z_0)$ is the Gouy phase and $\text{L}^{\lvert l\rvert}_p(\xi)$ are the generalized Laguerre polynomials,
\begin{equation}
\text{L}^{\lvert l\rvert}_p(\xi)=\sum_{m=0}^{p}(-1)^m\frac{(\lvert l\rvert+p)!}{(p-m)!\ (\lvert l\rvert+m)!\ m!}\xi^m.
\end{equation}
The two independent indices $l=0,\ \pm1,\ \pm2,\ ...$ and $p=0,\ 1,\ 2,\ ...$, correspond to the topological charge and the number of non-axial radial nodes of the mode. Note for $l=p=0$, we recover the standard Gaussian beam. Hence, the vector potential can be cast in the form,
\begin{equation}
\bm{E}(\bm{r},\ z,\ t)=\frac{1}{2}\bm{E_0}\ \text{LG}_{l,p}(r,\ \theta,\ z)e^{\text{i}w_0t-\text{i}k_0z}+c.c.,
\end{equation}
$\omega_0=ck_0$ is the frequency of laser, and $\bm{E_0}$ is the electric amplitude of the laser. Note that the wave-front is helical, in the Coulomb gauge, the electric can have a polarization component in the propagation direction. A realistic solution should include this longitudinal polarization component, which becomes particularly relevant near the vortex axis. However, to first order, both components will give independent dynamics, allowing us to consider only the transverse part.  

\begin{figure}
\includegraphics[width=8.5cm]{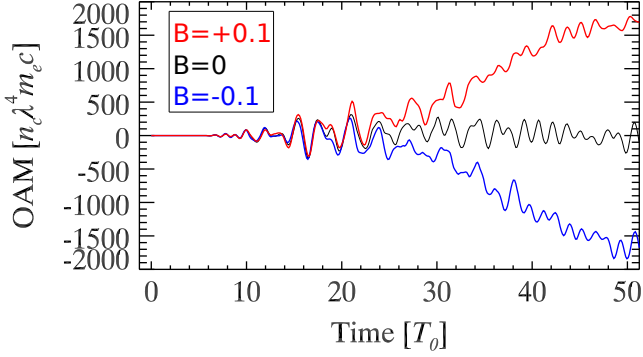}
\caption{\label{fig2} (color online) Total orbital angular momentum of plasmas as functions of time in 3D PIC simulations. Simulation parameters: normalized amplitude of twisted laser is of $a=5$, normalized plasma density is of $n_e=0.5$ and magetostatic seed of $B=0.1$ [red line (a)], $B=0$ [black line (b)] or $B=-0.1$ [blue line (c)] is loaded along z direction.}
\end{figure}

In the paraxial regime, LG modes carry a discrete OAM of $l\hbar$ per unit along their propagation directions. 
For a relativistic intense twisted-laser, the carried total OAM is huge. When interacting with plasmas, the total OAM absorbed by plasmas can be calculated by summarizing all individual-electrons. Here, as shown in Fig.\ 1, 
a linearly polarized LG laser with $\bm{E_0}=E_0\textbf{e}_x$ irradiates into uniform plasmas.
The laser electric field experienced by an individual electron is $E_x/E_0=(\lvert l\rvert !)^{-1/2}\times({r}/{b})^{\lvert l\rvert}\times\exp{(-r^2/b^2)}\times\cos(\l \theta+\omega^{\prime}t)$, with $p=0$. The x-direction velocity of that electron can be found as $u_x/u_{0}=(\lvert l\rvert !)^{-1/2}\times({r}/{b})^{\lvert l\rvert}\times\exp{(-r^2/b^2)}\times\sin(\l \theta+\omega^{\prime}t)$, with $u_{0}=eE_0/m_e\omega^{\prime}c$, $\omega^{\prime}=\omega_0(1-v_z/v_{\text{ph}})$, $v_z$ is the electron longitudinal velocity and $v_{\text{ph}}$ is the phase velocity of laser. Assume that the plasma is of uniform, the total OAM of the plasma system can be calculated as
\begin{equation}
\label{pz1}
P_z(t)=\int_0^t r\sin(\theta)\times Ln_em_eu_x \times r dr d\theta dt,
\end{equation}
where $L$ is the thickness of plasmas, $n_e$ is plasma density. After integrating over $r$ and $\theta$, we find that
\begin{equation}
\label{pz2}
\frac{P_z(t)}{P_0}=\left\{
\begin{aligned}
&& \pm \int_0^t \cos(\omega^{\prime}t) dt;\ l=\pm 1 \\
&& 0; \ l=0,\ \pm2,\ \pm3,\ ... \\
\end{aligned}
\right.
\end{equation}
From Eq.\ (\ref{pz1}) and (\ref{pz2}), we can see in the considered uniform plasmas cases, the twisted-light could transfer the carried OAM to plasmas only when the topological charge is of $l=\pm1$. However even if $l=\pm1$, the acquired total OAM is still zero after averaging with time, which means when laser penetrates through plasmas, the remained OAM in plasmas is zero. 

Here, we realize that it is possible for plasmas to acquire a net OAM, when there exist an external longitudinal magetostatic field. Although the magetostatic field is small, the barrier of transferring OAM from laser to plasmas can be broken.    

The motion of electron by considering an external magetostatic field ($\bm{B}=B_0\textbf{e}_z$) can be written as, $u_x/u_{0}=(\lvert l\rvert !)^{-1/2}\times({r}/{b})^{\lvert l\rvert}\times\exp{(-r^2/b^2)}\times\sin(\l \theta+\omega^{\prime}t)\times\cos(\omega_ct)$ and $u_y/u_{0}=(\lvert l\rvert !)^{-1/2}\times({r}/{b})^{\lvert l\rvert}\times\exp{(-r^2/b^2)}\times\sin(\l \theta+\omega^{\prime}t)\times\sin(\omega_ct)$, where $\omega_c=eB/m_e$ is Larmor frequency, which can be negative or positive depending on the direction of $B$. Similarly, the total OAM of plasmas can be written as
\begin{equation}
\label{pz3}
\frac{P_z(t)}{P_0}=\left\{
\begin{aligned}
\pm \int_0^t \cos[(\omega^{\prime}-\omega_c)t] dt;\ l=\pm1,\ \omega_c>0, \\
\pm \int_0^t \cos[(\omega^{\prime}+\omega_c)t] dt;\ l=\pm1,\ \omega_c<0. \\
\end{aligned}
\right.
\end{equation}
From Eq.\ (\ref{pz3}), we can see when $\omega^{\prime}=\omega_c$, plasma will continuously acquire OAM from lasers. This condition can be easily satisfied if the twisted-light is of relativistic intensity, whose ponderomotive force will always accelerate electrons forward, approaching light speed. Hence the experienced laser frequency by electrons will decrease with longitudinal velocity until the condition $\omega_0(1-v_z/v_{\text{ph}})=\omega_c$ is satisfied. When $\omega_0(1-v_z/v_{\text{ph}})=\omega_c$, we have $\cos[(\omega^{\prime}-\omega_c)t]=1$, plasmas will absorb OAM from twisted light with the value linearly increasing with time. Once plasmas acquire net OAM, the imposed magetostatic field would be further enhanced and hence electrons with smaller $v_z$ would also involve in. The magetostatic field and OAM in plasmas will repeatedly enhance with each other until $\omega_c=\omega_0$.

In detail, to trigger the magetostatic amplifier, we need to ensure that $\omega_0(1-v_z/v_{\text{ph}})=\omega_c$. 
A higher $v_z$ would result in a smaller demanding of magetostatic seed. Basically, by considering only the ponderomotive force of laser pulse, the maximal $v_z$ of electrons is increasing with laser amplitude $a$, i.e., $v_{z,\max}=a^2/2\gamma_{\max}$, with $\gamma_{\max}=1+a^2/2$. Considering the phase velocity of laser beam in plasmas is of $v_{\text{ph}}=1/\sqrt{1-\omega^2_{pe}/\omega^2_0\gamma}$, in the low density limit, the smallest demanding of laser amplitude is $a_{\min}=\sqrt{2(1-{B})/{B}}$, with ${B}=\omega_c/\omega_0$.

\textit{\textbf{Simulation results}}--In order confirm the proposed scheme, we have performed 3D PIC simulations by using LAPINE code. The simulation box is $20\ \mu\text{m}\ (x)\times20\ \mu\text{m}\ (y)\times20\ \mu\text{m}\ (z)$, which is divided into $200\times200\times400$ grids. A laser beam of LG mode ($l=1$ and $p=0$) is launched into simulation box at left boundary, with wavelength of $\lambda_0=1\ \mu\text{m}$, beam radius of $b_0=3\ \mu\text{m}$, beam duration of $5-20-5T_0$ ($T_0=\lambda_0/c$) and normalized amplitude $a=5$. Here, magetostatic seed ($B=+0.1$ or $B=-0.1$) is loaded along laser propagation direction, therefore the corresponding smallest demanding of laser amplitude is $a_{\min}=\sqrt{2(1-{B})/{B}}=4.24$. Uniform plasmas with normalized density $n_e=0.5n_c$ (the corresponding critical density is $n_c=1.1\times10^{21}\ /\text{cm}^3$ for laser of wavelength $1\ \mu\text{m}$) is placed between $z=5\ \mu\text{m}$ and $z=15\ \mu\text{m}$, with each cell containing $8$ electrons. The movement of ions is deliberately turned off as its contribution is ignitable for the considered situations.  

\begin{figure}
\includegraphics[width=8.5cm]{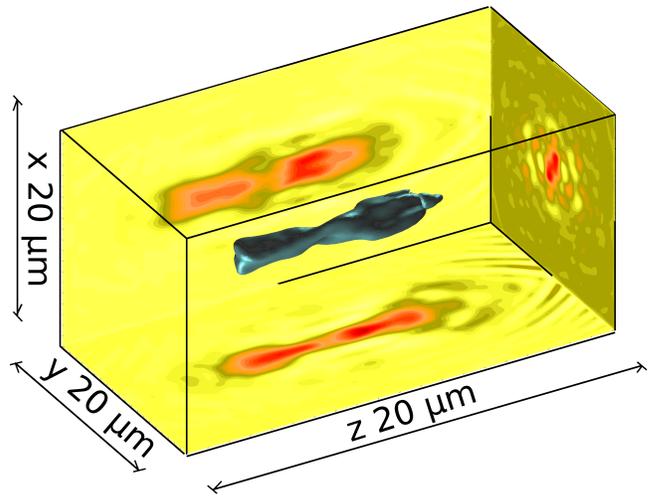}
\caption{\label{fig3} (color online) Isosurface visualisation of $B_z$ with data from 3D PIC simulations when the laser fully penetrating through plasmas. The x-y, x-z and y-z cross-sections of $E_x$ are also presented at the corresponding walls. Simulation parameters are the same as shown in Fig.\ \ref{fig1} but only with magetostatic seed of $B=0.1$ loaded along z direction.}
\end{figure}

\begin{figure}
\includegraphics[width=8.5cm]{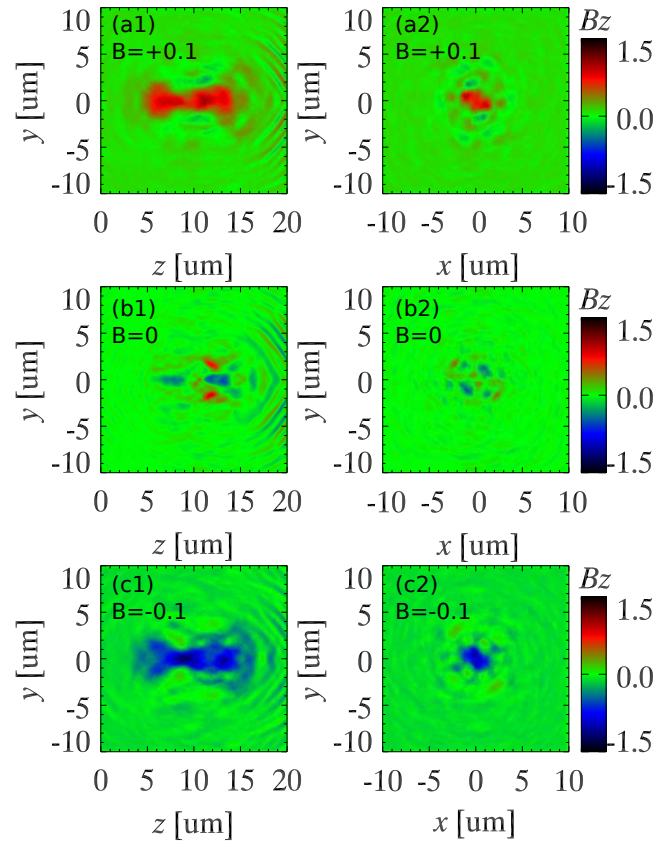}
\caption{\label{fig4} (color online) The y-z and x-y cross-sections of $B_z$ are presented in first and second columns. (a) (b) and (c) correspond to cases with different magetostatic seed of $B=0.1$ [red line (a)], $B=0$ [black line (b)] and $B=-0.1$ [blue line (c)] loaded long z direction. Simulation parameters are the same as shown in Fig.\ \ref{fig2}.}
\end{figure}

Fig.\ \ref{fig2} shows the total OAM of plasmas by summarizing all individual electrons as function of time. Here, black line is the case without magetostatic seed, we can see that within the laser-plasma interaction window ($5T_0<t<45T_0$), total OAM of plasmas just simply oscillates with time. After the penetration of laser beam, the remained OAM is close to zero. This kind of behaviour can be well described by Eq.\ (\ref{pz2}), i.e. $P_z(t)/P_0=\int_0^t\cos(\omega^{\prime}t)$. Red line and blue line are the cases with initial non-zero magetostatic seed, $B=+0.1$ for red line and $B=-0.1$ for blue line. We can see the total OAM of plasmas is continuously increasing/decreasing with time for magetostatic seed of $B=+0.1$/$B=-0.1$. Even after the penetration, the remained OAM in plasmas is still very large and kept constant. 

The huge OAM remained in plasmas would generate strong static magnetic fields. Here we have picked up the specific case with initial magetostatic seed of $B=+0.1$. As shown in Fig.\ \ref{fig3}, the isosurface of constant $B_z=0.5$ is plotted, and the x-y, x-z and y-z cross-sections of $B_z$ are also presented at the corresponding walls. The static magnetic fields produced in plasmas are of space-uniform and extremely intense. Detailed static magnetic fields profile are presented in Fig.\ \ref{fig4}. The first column corresponds to y-z cross-sections and the second column x-y cross-sections. We can see, for initial magetostatic seed of $B=+0.1$ [Fig.\ \ref{fig4} (a)], the finial magnitude of $B_z$ can be as high as $B=1$, i.e., $10000$ T considering the laser wavelength is of $1\ \mu\text{m}$. While for initial magetostatic seed of $B=-0.1$ [Fig.\ \ref{fig4} (c)], the final magnitude is $B=-1$. When no magetostatic seed is loaded initially, the finial $B_z$ is week and disturbed, as shown in Fig.\ \ref{fig4} (b). 

The upper limit of the proposed magetostatic amplifier is determined when the corresponding Larmor frequency is comparable to laser frequency in vacuum, i.e., $\omega_c=\omega_0$. For laser of wavelength $1\ \mu\text{m}$, after the amplification, the finial static magnetic field can reach $10000$ T. To obtain even higher static magnetic field, such as $20000$ T, $30000$ T, ...,  second-, third- or higher order harmonics of twisted-light shall be used.

\textit{\textbf{Discussion and Summary}}--We also noticed that strong magnetostatic fields can be produced by using circularly polarized Gaussian laser, through the so-called ``self-matching mechanism'' \cite{he1} in near critical density plasmas. To achieve ``self-matching mechanism'', laser intensity is usually of extremely strong, at $10^{21}\sim10^{22}\ \text{W}/\text{cm}^2$. Under such conditions, non-linear effects, such as laser self-focusing and helical electron bunch formations \cite{he2, he3} would become the dominant contribution. However our theory works in first order approximation, therefore it is a robust and well-controllable scheme.  

In summary, we have investigated relativistic intense twisted-light plasma interaction basically in the first-order approximation. To first order, plasma is assumed of uniform, the longitudinal component of laser field is ignored when compared with the transverse parts. Furthermore, instabilities, such as filament instabilities, are ignored in the considered situations. We find that, when twisted light interacts with plasmas, the net OAM acquired by plasmas is basically zero after interaction. However, we also find when imposing a small magetostatic seed along laser propagation direction, the barrier would be broken, giving rise to dramatic OAM transfer from LG lasers to plasmas. Hence, the net OAM remained in plasmas system would continuously enhance the magetostatic field, until the Larmor frequency of electrons ($\omega_c=eB/m_e$) is comparable to the laser frequency ($\omega_0$) in vacuum. Therefore, the upper limit of static magnetic field can be controlled by using higher order harmonics of twisted-light.
3D-PIC simulations are performed to confirm our theory, producing space-uniform, time-stable and extremely intense magetostatic fields.

\begin{acknowledgments}
This work was partially supported by National Natural Science Foundation of China (No. 11605269, 11674341 and 11675245). 
\end{acknowledgments}

{}

\end{document}